\documentclass[galaxies,article,accept,moreauthors,pdftex]{mdpi} 

\usepackage{graphics,epsf}
\usepackage{amsmath}                
\usepackage{amsfonts}               
\usepackage{amssymb}                
\usepackage{epsfig}                 
\usepackage{graphicx}

\def \kms{~\rm{km~s^{-1}}}

\def \msyr{~\rm{M_{\odot}}~\rm{yr^{-1}}}

\def \K{~\rm{K}}

\def \erg{~\rm{erg}}

\def \pc{~\rm{pc}}

\def \days{~\rm{days}}

\def \sun{\odot}
\def \rmModot{~\rm{M_{\sun}}}
\def \rmRodot{~\rm{R_{\sun}}}
\def \rmLodot{~\rm{L_{\sun}}}

\def \rmME{~\rm{M_{\oplus}}}
\def \rmRE{~\rm{R_{\oplus}}}

\makeatletter

\let\jnl@style=\rmfamily 
\def\ref@jnl#1{{\jnl@style#1}}%
\newcommand\aj{\ref@jnl{Astron. J.}}
\newcommand\araa{\ref@jnl{Annu. Rev. Astron. Astrophys.}}
\newcommand\apj{\ref@jnl{Astrophys. J.}}
\newcommand\apjl{\ref@jnl{Astrophys. J. Lett.}}     
\newcommand\apjs{\ref@jnl{Astrophys. J. Suppl.}}
\newcommand\ao{\ref@jnl{Appl. Opt.}}
\newcommand\apss{\ref@jnl{Astrophys. Space Sci.}}
\newcommand\aap{\ref@jnl{Astron. Astrophys.}}
\newcommand\aapr{\ref@jnl{Astron. Astrophys. Rev.}}
\newcommand\aaps{\ref@jnl{Astron. Astrophys. Suppl.}}
\newcommand\azh{\ref@jnl{AZh}}
\newcommand\baas{\ref@jnl{BAAS}}
\newcommand\icarus{\ref@jnl{Icarus}}
\newcommand\jaavso{\ref@jnl{JAAVSO}}  
\newcommand\jrasc{\ref@jnl{JRASC}}
\newcommand\memras{\ref@jnl{MmRAS}}
\newcommand\mnras{\ref@jnl{Mon. Not. R. Astron. Soc.}}
\newcommand\pra{\ref@jnl{PhRvA}}
\newcommand\prb{\ref@jnl{PhRvB}}
\newcommand\prc{\ref@jnl{PhRvC}}
\newcommand\prd{\ref@jnl{PhRvD}}
\newcommand\pre{\ref@jnl{PhRvE}}
\newcommand\prl{\ref@jnl{PhRvL}}
\newcommand\pasp{\ref@jnl{PASP}}
\newcommand\pasj{\ref@jnl{PASJ}}
\newcommand\qjras{\ref@jnl{QJRAS}}
\newcommand\skytel{\ref@jnl{S\&T}}
\newcommand\solphys{\ref@jnl{SoPh}}
\newcommand\sovast{\ref@jnl{Soviet~Ast.}}
\newcommand\ssr{\ref@jnl{SSRv}}
\newcommand\zap{\ref@jnl{ZA}}
\newcommand\nat{\ref@jnl{Nature}}
\newcommand\iaucirc{\ref@jnl{IAUC}}
\newcommand\aplett{\ref@jnl{Astrophys.~Lett.}}
\newcommand\apspr{\ref@jnl{Astrophys.~Space~Phys.~Res.}}
\newcommand\bain{\ref@jnl{BAN}}
\newcommand\fcp{\ref@jnl{FCPh}}
\newcommand\gca{\ref@jnl{GeoCoA}}
\newcommand\grl{\ref@jnl{Geophys.~Res.~Lett.}}
\newcommand\jcp{\ref@jnl{JChPh}}
\newcommand\jgr{\ref@jnl{J.~Geophys.~Res.}}
\newcommand\jqsrt{\ref@jnl{JQSRT}}
\newcommand\memsai{\ref@jnl{MmSAI}}
\newcommand\nphysa{\ref@jnl{NuPhA}}
\newcommand\physrep{\ref@jnl{PhR}}
\newcommand\physscr{\ref@jnl{PhyS}}
\newcommand\planss{\ref@jnl{Planet.~Space~Sci.}}
\newcommand\procspie{\ref@jnl{Proc.~SPIE}}

\newcommand\actaa{\ref@jnl{AcA}}
\newcommand\caa{\ref@jnl{ChA\&A}}
\newcommand\cjaa{\ref@jnl{ChJA\&A}}
\newcommand\jcap{\ref@jnl{JCAP}}
\newcommand\na{\ref@jnl{NewA}}
\newcommand\nar{\ref@jnl{NewAR}}
\newcommand\pasa{\ref@jnl{PASA}}
\newcommand\rmxaa{\ref@jnl{RMxAA}}

\newcommand\maps{\ref@jnl{M\&PS}}
\newcommand\aas{\ref@jnl{AAS Meeting Abstracts}}
\newcommand\dps{\ref@jnl{AAS/DPS Meeting Abstracts}}

\makeatother

\firstpage{1} 
\pubvolume{xx}
\issuenum{1}
\articlenumber{5}
\pubyear{2020}
\copyrightyear{2020}
\history{Received: date; Accepted: date; Published: date}
\updates{yes} 

\Title{ASASSN-13db 2014--2017 Eruption as~an~Intermediate Luminosity Optical Transient}

\Author{Amit Kashi *\orcidA{}, Amir M. Michaelis\orcidB{} and~Leon Feigin\orcidC{}} 

\address[1]{
 Department of Physics, Ariel~University, P.O. Box  3, Ariel~4070000,~Israel 
}

\corres{\hangafter=1 \hangindent=1.0em \hspace{-1em} Correspondence: kashi@ariel.ac.il}

\abstract{
The~low mass star ASASSN-13db experienced an~EXor outburst in 2013, which~identified it as~a~Young Stellar Object (YSO). Then, from 2014 to 2017 it had another outburst, longer and~more luminous than the~earlier. We~analyze the~observations of the~second outburst, and~compare it to eruptions of Intermediate Luminosity Optical Transients (ILOTs). We~show that the~decline of the~light curve is almost identical to that of the~V838~Mon, a~prototype of a~type of ILOT known as~Luminous Red Nova (LRN). This~similarity becomes conspicuous when oscillations that are associated with rotation are filtered out from the~light curve of ASASSN-13db. We~suggest that the~eruption was the~result of accretion of a~proto-planet of a~few Earth masses. The~proto-planet was shredded by tidal forces before it was accreted onto the~YSO, releasing gravitational energy that powered the~outburst for $\approx 800 \days$, and~ended in a~$\approx 55 \days$ decline phase. When the~accretion material started depleting the~accretion rate lowered and~the~eruption light curve declined for almost two months. Then it exhausted completely, creating a~sharp break in the~light curve. Another possibility is that the~mass was a~result of an~instability in the~proto-planetary disk that lead to a~large episode of accretion from an~inner viscous disk. We~find that the~variation of the~temperature of the~outburst is consistent with the~surface temperature expected from a~depleted viscous accretion disk. The~2014-2017 outburst of ASASSN-13db may be the~least energetic ILOT to have been discovered to date, with an~energy budget of only $\approx 10^{42} \erg$.
}

\keyword{planet-star interactions ; accretion, accretion discs ; stars: pre-main-sequence ; (stars:) binaries: general}

\begin{document}


\section{Introduction}
\label{sec:intro}

Intermediate luminosity optical transients (ILOTs) are exotic outbursts with luminosities which~fall between those of novae and~supernovae (SN). Many new ILOTs are being discovered by modern surveys and~dedicated campaigns (e.g.,~\cite{Mouldetal1990, Bondetal2003, Rauetal2007, Rauetal2009, Ofeketal2008, Ofeketal2016, Prietoetal2009, Botticella2009, Smithetal2009, Berger2009a, Berger2009b, KulkarniKasliwal2009, Mason2010, Pastorello2010,Pastorelloetal2019a, Pastorelloetal2019b, Kasliwaletal2011, Kasliwal2013, Tylendaetal2013, Kurtenkovetal2015, Smarttetal2015, Williamsetal2015, Ofeketal2016, Pejchaetal2016a, Pejchaetal2016b, Tartagliaetal2016, Villaretal2016, Humphreysetal2017, Humphreysetal2019, Blagorodnovaetal2017, Adamsetal2018, Bellmetal2019, Jayasingheetal2019}. 
 The~group consists of many different astronomical eruptions which~are diverse, but~are found to have shared properties~\citep{Kashietal2010, SokerKashi2016, Kashi2018Galaxies}. Kashi and~collaborators~\cite{Kashietal2010} noticed similarities in the~light curves of a~few ILOTs, that were not considered to be~similar. The~shape of the~light curve may start with a~few peaks (frequently~two~peaks) followed by a~downward concave curve. At~first sight, the~time-scale for decline, and~the~absolute magnitude of the~curve may look different. Their~similarity becomes evident when they are re-scaled, shifting the~peak luminosities to coincide just before the~decline in the~light curve, and~multiplying the~light curves by a~scaling factor.
In \cite{Kashietal2010} it was proposed that this similarity could indicate that a~common physical mechanism is involved, that has a~fingerprint in the~form of this similar decline.

The~Energy-Time Diagram (ETD; \cite{Kashietal2010}, ~\cite{Kashi2018Galaxies})
 is used for classifying transients in terms of their characteristic time of decline and~their total energy (The~ETD webpage: \url{phsites.technion.ac.il/soker/ilot-club/}.). 
 In~the~ETD many of the~transients form a~slated stripe more energetic, for~a~given time scale, than novae yet less energetic than SNe. We~refer to this stripe as~the~Optical Transient Stripe. In~\cite{Kashietal2010} it was suggested that most objects that populate the~stripe share a~similar powering mechanism, that they suggest to be accretion energy. The~different time scale are related to the~accretion rate or the~mass supply rate, that in some cases can be limited by the~time it takes to dissipate the~angular momentum of the~accreted mass through~viscosity.

One well studied transient is the~2002 outburst of V838~Mon~\citep{Bondetal2003}. It~is considered a~prototype of a~sub-type of ILOTs, known as~Luminous Red Novae (LRN). Both classical nova and~He-shell flash were suggested as~explanations for the~unusual eruption, but~later ruled out~\citep{TylendaSoker2006}. The~star involved in the~V838 Mon eruption was a~massive star, possibly even on the~main-sequence (MS). Perhaps~the~most unusual observation was that as~time passed V838~Mon became redder\citep{Evansetal2003,Starrfieldetal2005}, which~is exactly opposite to the~evolution of classical~novae.

A~model for the~eruption of V838~Mon that involves a~stellar merger event followed by an~accretion process, in which~the~surviving star accreted the~material of its destructed companion, was proposed by~\cite{SokerTylenda2006}. The~model was named the~merger-burst model. This~model was able to account for all observations and~its only drawback was that its parameters were not strictly constrained (see~also~\cite{Tylendaetal2005, TylendaSoker2006}). The~binary merger scenario in V838~Mon was strengthened by the~eruption of the~LRN V1309~Sco~\citep{Mason2010} where a~clear signal of an~eclipsing binary was observed prior to the~eruption~\citep{Tylendaetal2011}.

Recently,~\cite{Pastorelloetal2019b} also considered both single star and~binary models for a~collection of LRNe, and~favored a~binary merger model with common envelope ejection. They~made a~distinction for objects fainter than $M_V=-10$, which~they termed Red Novae, and~brighter than this value (typically~at~$M_V=-12$--$-15$), which~they termed Luminous Red Novae. They~suggested that different stellar progenitors are responsible for each of the~two, but~agreed that LRNe are ``scaled-up'' Red Novae. As~we are interested in the~physical mechanism we will not use this~subdivision.

Interestingly,~\cite{Bearetal2011} suggested that merger of a~planet with a~Brown Dwarf or low mass star can lead to an~eruption of shorter time scale and~slightly less energetic, that would be on the~lower left side of the~optical transient stripe. It~later was discovered that some eruptions can have shared properties with ILOTs even if they are external to the~optical transient stripe~\citep{KashiSoker2017, Soker2018}. In~\cite{KashiSoker2017} it was
suggested that the~unusual outburst of the~Young Stellar Object (YSO) ASASSN-15qi~\citep{Herczegetal2016} is an~ILOT event, similar in many respects to LRN events such~as V838~Mon, but~much fainter and~of lower total energy. In~other words, ASASSN-15qi is an~unusual ILOT in the~sense that it has low power and~therefore resides below the~OTS. The~physical model for ASASSN-15qi suggests that in a~similar manner to LRNe, a~secondary object was tidally destroyed onto the~primary YSO, releasing gravitational energy in the~process~\citep{KashiSoker2017}. They~suggested that the~secondary object was a~Saturn-like \textit{planet} instead of a~low mass pre-MS companion in the~LRN model. This~process created an~accretion disk and~manifested as~a~gravitationally powered ILOT. Differently from V838~Mon which~is much more energetic, the~mass of the~destructed planet is too low to cause the~YSO to have an~inflated envelope, and~hence the~merger remnant stays hot, and~therefore does not redden as~the~LRNe.

In~addition to ASASSN-15qi,~\cite{Kashi2018Galaxies} further suggested that another transient, ASASSN-13db~\cite{SiciliaAguilaretal2017}, may also be a~related object. The~variable low-mass star ASASSN-13db was identified by the All Sky Automated Survey for SuperNovae (ASAS-SN; e.g., \cite{Jayasingheetal2019}), after brightening in the~visual by $\approx 4$~mag in September~2013
(\cite{Holoienetal2014, Prietoetal2014, Shappeeetal2014}). The~outburst occurred in an~M5 young star surrounded with a~proto-planetary~disk. FU~Ori outbursts (e.g.,~\cite{HartmannKenyon1996,Audardetal2014} and~references~therein) occur in pre-MS stars, and~are observed as~an~extreme
 change in magnitude with a~slow decline that may last for years. The~spectral type also changes during the~outburst. FU~Ori outbursts can be regarded as~more energetic counterparts of the~EXor class of outburst, of which~EX~Lupis is a~prototype~\citep{Herbig2007}. These are pre-MS variables that show flares of a~few months to a~few years, and~of several magnitudes amplitude, as~a~result of episodic mass accretion~\citep{Herbig1998, Audardetal2014}. The~various intensities and~various timescales are not only found observationally, but~are also expected from theory~\citep{Clarkeetal2005, Zhuetal2009, Vorobyovetal2013}. ASASSN-13db was classified as~an~EXor but~with remarkable luminosity, comparable to EX Lupi 2008 outburst~\citep{Aspinetal2010}, so that it may be a~link between FUors and~EXors~\citep{ContrerasPenaetal2017}.

\textls[-15]{The~2013 outburst was not the~last word heard from ASASSN-13db~\citep{SiciliaAguilaretal2017}. Later, in 2014, ASASSN-13db had another outburst of a~different nature (not an~FU~Ori outburst), in~which~the~peak magnitude in the~$V$-band was brighter by about $1$~mag than the~outburst in~2013. The~entire 2014--2017 transient lasted about~800~days. It~stayed bright at~a~high plateau for about~500~days, experiencing fluctuations of $\delta V \simeq 1$~mag. Later it started gradually declining for another $\approx 250 \days$, which~ended in~2017 with a~fast decline by approximately~2.5 magnitudes over two months (see~Figure~1 of ~\cite{SiciliaAguilaretal2017}).}

\textls[-5]{In~this paper we examine the~possibility that the~2014--2017 outburst of ASASSN-13db (hereafter~A13db1417) was~an~ILOT similar to a~LRN. In~Section~\ref{sec:obs} we discuss its properties in more detail, and~compare it to a~LRN. In~Section~\ref{sec:model} we propose a~model for this outburst. Our~summary and~discussion appear in Section~\ref{sec:summary}.}

\section{Observational Comparison}
\label{sec:obs}

According to~\cite{Holoienetal2014} ASASSN-13db is an~M5 star with mass $M_1 \sim 0.15 \rmModot$, luminosity $L_1 \simeq 0.06 \rmLodot$ and~radius $R_1 \sim 1.1 \rmRodot$. These values were measured during the~short quiescence phase after the~2013 outburst.
The~distance of ASASSN-13db is estimated to be $\approx 380 \pc$~\citep{SiciliaAguilaretal2017}. We~will adopt these properties in our analysis.

\begin{figure}[t]
\centering
\begin{tabular}{c}
\includegraphics[trim= 0.9cm 1.5cm 1.0cm 0.8cm,clip=true,width=0.9\textwidth]{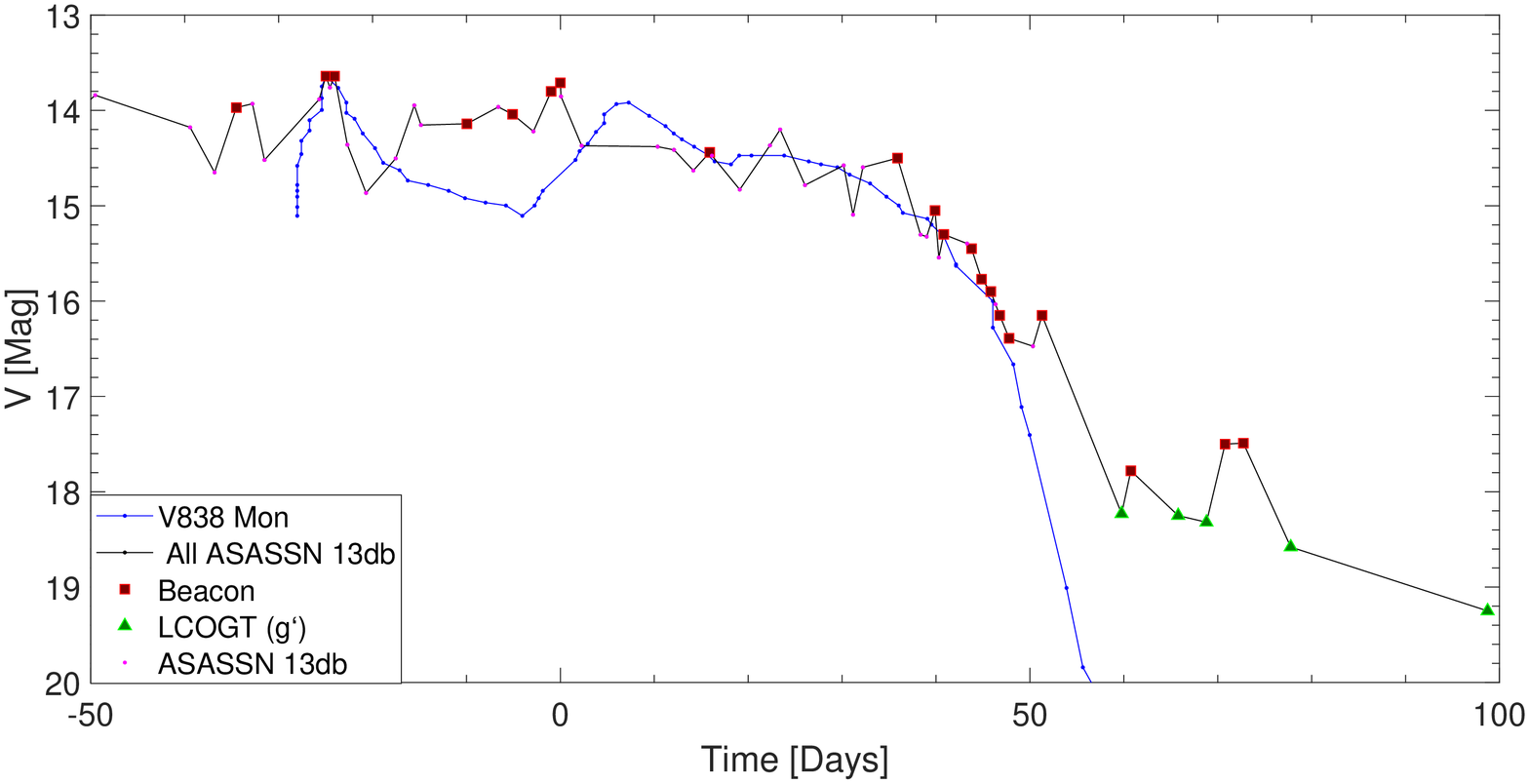}\\
(\textbf{a}) \\ 
\includegraphics[trim= 0.9cm 0.0cm 1.0cm 0.8cm,clip=true,width=0.9\textwidth]{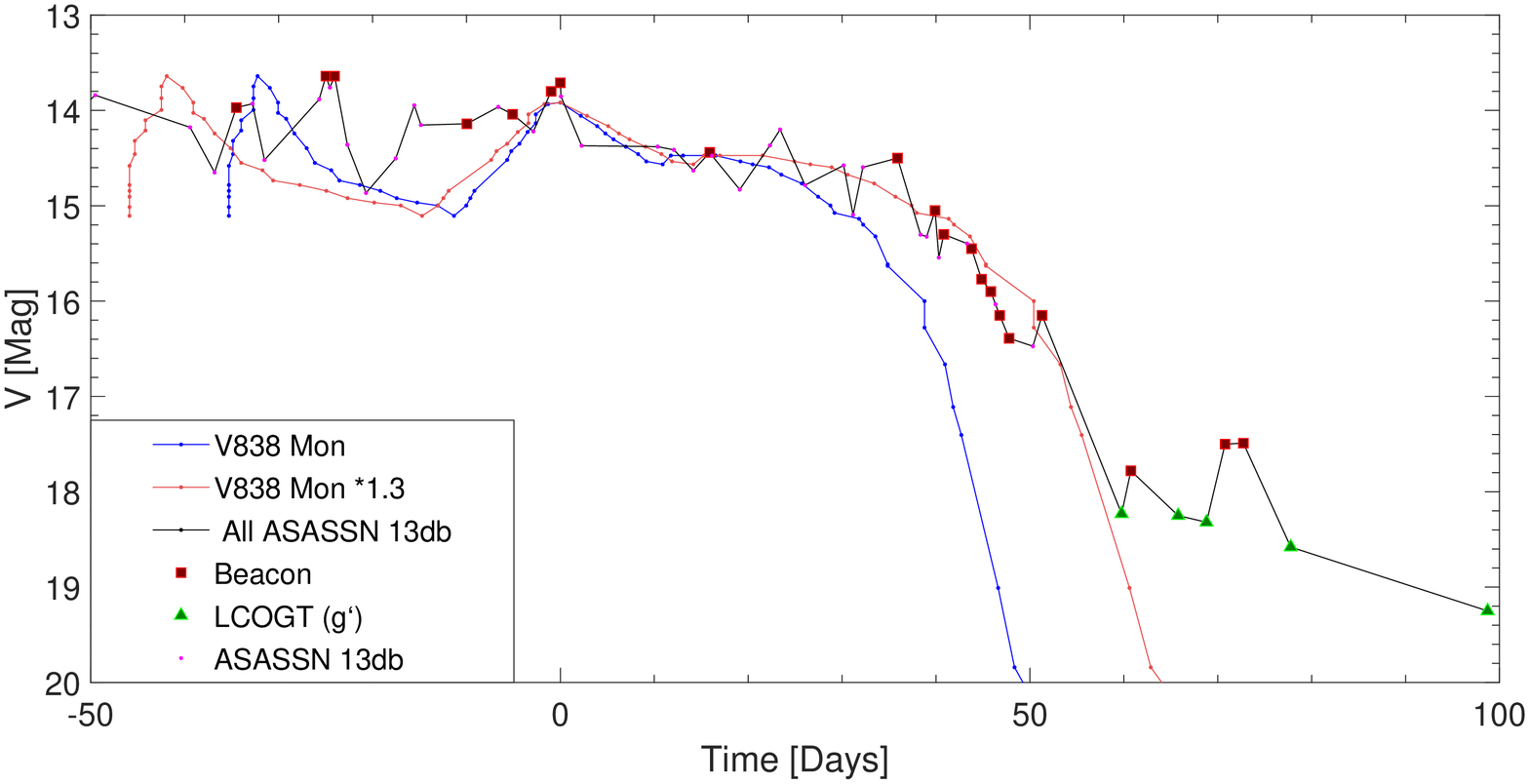}\\
(\textbf{b}) \\
\end{tabular}
\caption{\textls[-15](\textbf{a}) Comparing the~$V$-band light curves of A13db1417~\citep{SiciliaAguilaretal2017} and~V838~Mon (\cite{Bondetal2003, Starrfieldetal2005,Sparksetal2008}). The~magnitude scale is the~apparent magnitude for A13db1417. The~light curve of V838~Mon was shifted by $\Delta V_{\rm V838~Mon} =6.9$~mag to match the~second peak before decline. The~time axis  focuses on the~end of the~$\approx 800 \days$ duration of A13db1417 (see \cite{SiciliaAguilaretal2017}) which~is the~$\simeq 55 \days$ decline phase. The~peak at~$\rm{JD} \simeq 2457728$ marks $t=0$. (\textbf{b}) Same us the~upper panel, but~the light curves were shifted to match the~peak just before decline. In~addition, the~time axis of the~light curve V838~Mon is scaled by a~factor of 1.3 relative to the~matched peak. This~results in that the~two light curves match for about~4~mag. 
}
\label{fig:v838_VS_13db}
\end{figure}

We~focus on the~decline phase of A13db1417.
First, we gathered the~light curves from the~three observatories from where the~data was provided separately in~\citep{SiciliaAguilaretal2017}: ASASSN $V$-band, Beacon $V$-band, and~LCOGT $g'$-band, into one combined light curve. This~will appear later in Figure~\ref{fig:v838_VS_13db} as~``All~ASASSN~13db''. As~evident in the~observations of~\cite{SiciliaAguilaretal2017}, the~decline phase starts at~$V \simeq 14$ and~continues up to the~$V \simeq 18$, in a~generally concave shape. Then, $\approx51$--$59 \days$ after the~peak (there~is a~gap in the~observations that makes it hard to~tell), the~light curve breaks to a~shallow, bumpy~decline.

In~the~study of SNe, it is a~common practice to stretch the~light curves in order to match them (e.g.,~\cite{Goldhaberetal2001,Conleyetal2008}). This~practice is essentially what made possible to derive the~Parker law for SNe Ia, that~allows using them as~standard candles for determining cosmological distances. A~reassembling process was recently suggested for classical novae~\citep{HachisuKato2019}. This~process was used for identifying candidate ILOTs and~comparing between eruptions~\citep{Kashietal2010}, and~developing models for possible new ILOTs (e.g.,~\cite{Bearetal2011}).

To apply this procedure we shifted down the~much more luminous light curve of the~2002 outburst of V838~Mon~\citep{Bondetal2003,Starrfieldetal2005,Sparksetal2008} by $6.9$ mag, to have its peak at~the~same brightness as~that of A13db1417.
We~then superimposed the~light curve of the~outburst of V838~Mon on A13db1417.

Figure~\ref{fig:v838_VS_13db} shows in the~upper panel the~result of only a~simple shift, for which~the~second peak before the~decline (one before last) was matched for both transients.
The~two transients show a~similar decline for about 3 magnitudes.
The~light curves of the~two transients separate only after this shared decline, as~A13db1417 at~this point has a~bump in the~light curve (JD$\simeq2457779$, which~is $\simeq50 \days$ after the~last peak before decline).
On~the~lower panel, we matched the~peak just before the~decline starts. We~find that when the~time axis of the~light curve of V838~Mon is scaled by a~factor of 1.3 the~two light curves behave similarly for a~decline of $\Delta V \simeq 4$~mag.

A~period of $\sim$4.2 days superimposed in the~light curve of A13db1417 was found by~\cite{SiciliaAguilaretal2017}. Their~explanation for that periodicity, which~can even be seen by visually examining the~data, was~stellar rotation and~the~presence of spots. We~wish to filter out this effect from the~data, as~we believe that this effect contaminates a~light curve that is governed by accretion. We~note that we filter out the~variation regardless of its source, may it be spots, binary interaction or a~different~source.

The~A13db1417 signal includes high rate temporal sampling, sometimes a~few times per night and~on other times one measurement per a~few days. We~analyze the~unequally spaced signal power spectrum using the~generalized Lomb-Scargle periodogram (GLSP;~\cite{ZechmeisterKurster2009}), and~obtain a~wealth of high frequencies in the~power spectrum. We~use the~following method to transfer A13db1417 unevenly spaced data to a~per night regular sampled signal. We~first focus on the~data points that were sampled more than once at~a~night and~decimate the~signal using simple average result with a~per night data point. Then we use spline interpolation on the~signal to estimate the~missing data points. This~results in a~regular sampled signal per night. We~compare the~GLSP analysis with our new regular signal using Welch method~\citep{Welch1967} as~well as~the~GLSP (setting the~relevant parameters to evenly spaced sampling) power spectrum making sure we did not lose information (mainly more than 4 days periods).  As~a~final step we apply the~median filter (again~after checking with the~power spectrum and~making sure we do not lose information) to~emphasize the~signal trend behavior over the~fast temporal~fluctuation.

\begin{figure}[t]
\centering
\includegraphics[trim= 0.9cm 0.0cm 1.0cm 0.8cm,clip=true,width=0.9\textwidth]{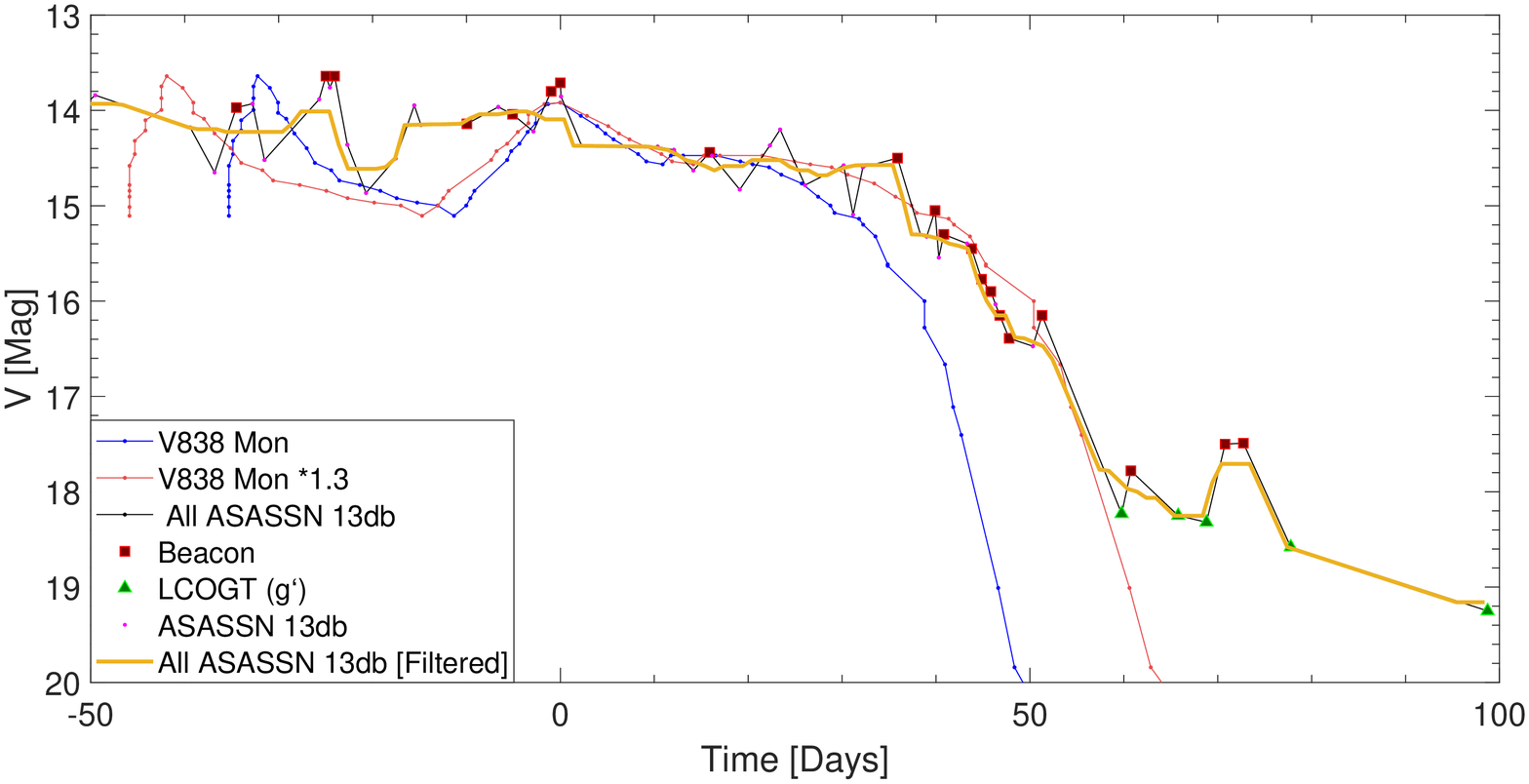} 
\caption{The~light curve of A13db1417, with the~variability resulted by rotation of stellar spot filtered out, compared to the~light curve of V838~Mon. This~variability caused oscillations of $\delta V \simeq 1$~mag. By~filtering out the~effect of rotation, we isolate the~component resulted from accretion.
The~filtered signal matches better the~scaled light curve of V838~Mon.}
\label{fig:v838_VS_13db_filtered}
\end{figure}

Figure~\ref{fig:v838_VS_13db_filtered} shows the~results of this process.
The~obtained light curve is much more refined and~less bumpy than the~original.
Therefore, much more resembling the~smooth decline light curve of V838~Mon.
We~can see~that as~expected the~two light curves match much~better.

\begin{figure}[t]
\centering
\includegraphics[trim= 0.9cm 0.0cm 1.0cm 0.8cm,clip=true,width=0.9\textwidth]{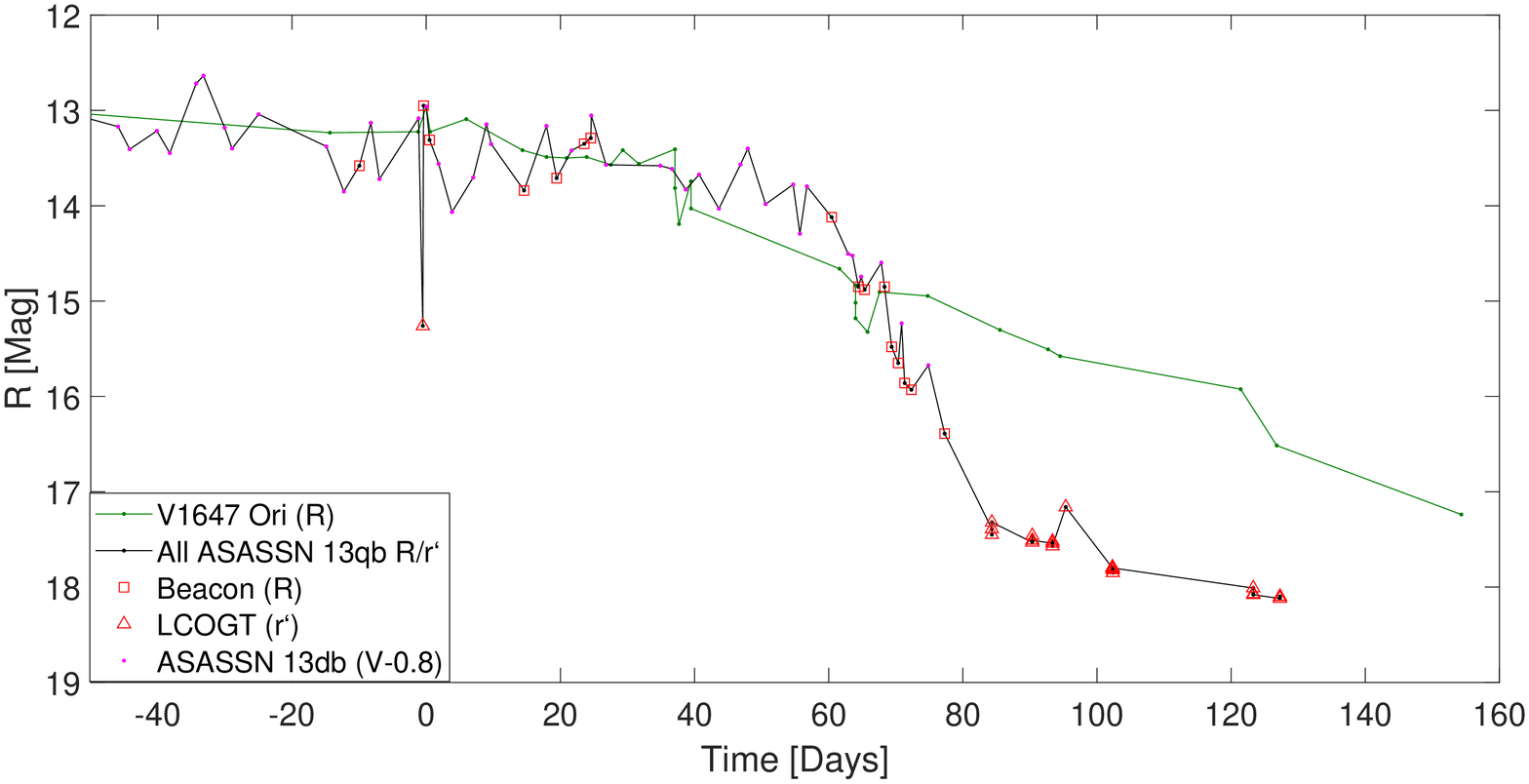} 
\caption{\textls[-12]{The~light curve of A13db1417 compared to that of the~FU~Ori eruption of V1647~Ori~\citep{Fedeleetal2007}. It~can be seen that the~two curves are very different and~the~decline does not follow the~same slope. This~suggest that this objects are different. Note that the~LCOGT observation at~$t=-2 \days$ may be an~outlier.}
}
\label{fig:V1647Ori}
\end{figure}

According to~\cite{SiciliaAguilaretal2017}, the~light curve of A13db1417 has an~overall shape that strongly resembles in length, duration, and~general shape (including~the~quick magnitude
drop after a~slow fading) the~light curve of the~FU~Ori variable V1647~Ori that erupted in 2004 (\cite{Fedeleetal2007}~and references therein). The authors of \cite{SiciliaAguilaretal2017} did not, however, present any comparison of these two light~curves. Reference~\cite{Fedeleetal2007} do not bring $V$-band observations of V1647~Ori, but~rather $R_c$-band covering well the~outburst and~its decline. In~Figure~\ref{fig:V1647Ori} we~compare the~$R$-band/$r'$-band light curve of A13db1417 with the~$R_c$-band of V1647~Ori. As~there are no ASASSN $R$-band observations, but~the~observations from Beacon and~LCOGT show thar $R-V$ and~$r'-g'$ are almost constant, we~adopt a~constant shift of $V-R=0.8$ for the~ASASSN observations, and~use its $V$-band observations to obtain the~$R$-band. We~evaluate the~error in this method is~$\pm 0.1$~mag.  Figure~\ref{fig:V1647Ori} shows that, though we try to make the~two light curves match, the~two objects have quite different decline slopes, with that of V1647~Ori being much steeper. The~comparison to V838~Mon is by far superior. We~consider it to be another hint that A13db1417 is not an~FU~Ori~eruption.

\begin{figure}[t]
\centering
\includegraphics[trim= 0.0cm 0.0cm 0.0cm 0.0cm,clip=true,width=0.9\textwidth]{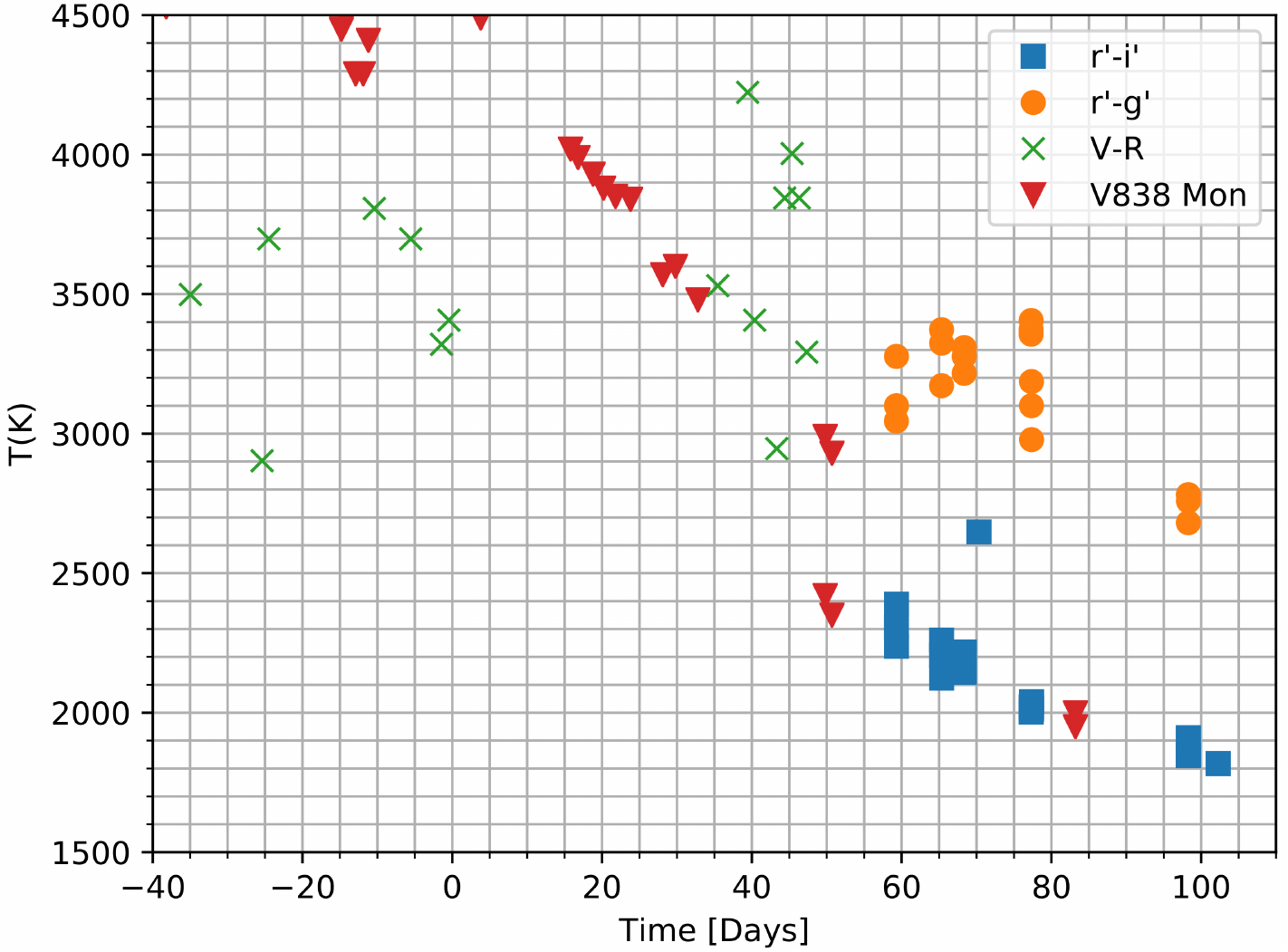} 
\caption{The~effective temperature of A13db1417, obtained from different filters as~indicated in the~legend. The~calculation was performed assuming black-body emission, which~is apparently not the~emitting spectrum, hence the~differences between the~estimated in different filter pairs.
Nevertheless we can see~that the~effective temperature is declining from $\simeq 4500\K$ to $\simeq 2000\K$ during the~eruption. Over-plotted is the~effective temperature from of V838~Mon, adopted from~\citep{Tylenda2005}. It~is evident that both objects have a~similar decline. }
\label{fig:Tt}
\end{figure}

We~calculate the~temperature from the~available observations, but~the~observations are quite noisy, and~when calculating color the~noise does not cancel. The~results during the~decline show that the~effective temperature slowly declines with time during the~55 days of disk dissipation (Figure~\ref{fig:Tt}). The~observations are insufficient for performing detailed comparison. However, we can compare them to the~evolution of the~effective temperature of V838~Mon (see~figure 2 in~\citep{Tylenda2005}). We~can see~that both objects had their effective temperature declining from $\simeq 4500\K$ to $\simeq 2000\K$ during the~eruption, a~similarity that adds to the~similarity in the~light curve we showed~above.

\section{An~ILOT Model}
\label{sec:model}

In~Section~\ref{sec:obs} we showed that the~characteristic decline of the~light curve of A13db1417 was similar to that of  V838~Mon. We~adopt the~model of~\cite{SokerTylenda2006} that suggests that a~merger event lead to the~outburst of V838~Mon. In~\cite{SokerTylenda2006} it was estimated that the~eruption of V838~Mon involved ejecting mass with kinetic energy of $E_{\rm kin} \approx 10^{44} \erg$, which~is about an~order of magnitude larger than the~emitted radiated energy of the~eruption. As~mentioned earlier in Section~\ref{sec:intro}, in~\cite{KashiSoker2017} the~properties of the~YSO eruption ASASSN-15qi were compared to those of V838~Mon. The~merger-like scenario that they suggested involved a~secondary object that was tidally destroyed onto the~primary $\approx 2.4 \rmModot$ YSO, releasing gravitational energy in the~process.
They~estimated that the~ejected mass could have a~similar value for the~kinetic energy, $E_{\rm kin} \approx 10^{44} \erg$.

We~now turn to examine a~model in which~A13db1417 was a~result of a~destruction of a~proto-planet onto the~YSO. The~innermost proto-planetary disk gas can be drained during the~previous EXor outburst, creating a~``disk gap'' (e.g.,~\cite{Gotoetal2011,BanzattiPontoppidan2015}). But~then in~2014 the~planet arrives and~shredded into fragments that from a~temporary disk, that exists until its depletion in~2017. Such~a~destruction would be the~result of instabilities in the~proto-planetary disk that may contain a~number of large bodies that interact with each other and~with the~material in the~proto-planetary disk. If~such an~object comes into the~tidal destruction radius of the~YSO, it~would be shredded and~form an~accretion disk. The~material would fall onto the~star, releasing gravitational energy. The~process would continue until the~available material is depleted, after~which~the~light curve would decline. It~is also possible, as~we discuss below, that the~instability in the~disk causes accretion of material in the~form of gas rather than a~destructed planet. It~is only required that the~arriving material would posses enough angular momentum to form a~close accretion disk around the~proto-star.

Modeling the~observations of A13db1417,~\cite{SiciliaAguilaretal2017} suggested accretion rates $\lesssim 3 \times 10^{-7} \msyr$. It~is not mentioned, however, whether this estimate includes the~kinetic energy (Namely, it was not mentioned whether the~supply of the~accreted gas can account only for radiation or for all the~energy). The~radiated energy during the~$\approx 800$ days of A13db1417 is $E_{\rm rad} \approx 2\times 10^{41} \erg$. The~wings of the~H$\alpha$ line observed in A13db1417 reach $300 \kms$~\cite{SiciliaAguilaretal2017}, indicating the~gas reach that velocity. We~will calibrate the~calculations of the~energy with $v_{\rm ej}=300 \kms$.
The~kinetic energy involved in the~eruption of A13db1417 can therefore be calibrated~as
\begin{equation}
E_{\rm kin} \simeq 9 \times 10^{41}
\left(\frac{M_{\rm ej}}{10^{-6}\rmModot}\right)
\left(\frac{v_{\rm ej}}{300 \kms}\right)^2 \erg, 
\label{eq:Ekin}
\end{equation}
where $M_{\rm ej}$ is the~ejected mass. A~lower velocity would result smaller energy, but~as~most of the~mass probably travels in the~range of $\gtrsim 100 \kms$, the~energy should still be in the~range of few $\times 10^{41} \erg$. As~the~accreted mass is small, we do not expect the~envelope to inflate, a~process that may consume more energy, which~is not observed. Namely the~total energy involved in the~A13db1417 eruption is $E_{\rm tot} \approx 10^{42} \erg$.
The~gravitational energy that can be supplied through accretion of a~total amount of mass $M_{\rm acc}$ from a~destructed object can be calibrated with the~parameters of the~YSO as
\begin{equation}
E_{\rm grav} \simeq 2.6 \times 10^{42}
\left(\frac{M_1}{0.15\rmModot}\right)
\left(\frac{M_{\rm acc}}{10^{-5}\rmModot}\right) 
\times \left(\frac{R_1}{1.1 \rmRodot}\right)^{-1} \erg. \\ 
\label{eq:Erad}
\end{equation}
Namely, the~accreted object has the~mass of a~planet, possibly a~proto-super-earth planet with mass of few $\times 10^{-6}$--$ 10^{-5}\rmModot$. The~average accretion rate we obtain over the~$t_{\rm acc} \approx 800 \days$ of the~eruption~is
\begin{equation}
\dot{M}_{\rm acc} \simeq 4 \times 10^{-6}
\left(\frac{M_p}{3 \rmME}\right) \left(\frac{t_{\rm acc}}{800 \days}\right)^{-1} \msyr. 
\label{eq:mdotacc}
\end{equation}
This~value (for the~accretion rate during the~outburst) is about 3000 times larger than the~quiescence accretion rate from the~proto-planetary disk, obtained from spectroscopic lines~\citep{SiciliaAguilaretal2017}. It~might be that the~mass is supplied not from a~proto-planet but~rather directly from the~proto-planetary disk, that~due~to an~instability increases the~accretion rate for the~duration of the~eruption. In~this case an~inner denser disk may from (or~a~thicker disk-like structure) that contains the~mass that was transferred from the~protoplantary disk to the~star, while it is being consumed by the~proto-star.

The~radii of proto-planets depend on their age, the~size of the~core, the~accretion rate onto the~core (if they are created by accretion of planetesimals), the~planet pressure profile, and~the~efficiency of cooling with the~presence (or~obscuration) of stellar irradiation and~tidal heating (e.g., \cite{Baruteauetal2016, Fortneyetal2007, Guillot2005} and~references~therein). According~to these studies, at~a~young age the~radius can be twice or more the~final radius. \cite{Fortneyetal2007} emphasize that planets that form closer to their star have a~larger radius, especially at~young ages.
Therefore the~expected density of a~proto-super-earth~is

\begin{equation}
\rho_p \simeq 1.06
\left(\frac{M_p}{3 \rmME}\right)
\left(\frac{R_p}{2.5 \rmRE}\right)^{-1}~\rm{g~cm^{-3}}, 
\label{eq:rho_p}
\end{equation}
where $M_p$ and~$R_p$ are the~planet mass and~radius, respectively.

We~would like to clarify that when we use the~name super-earth we do not necessarily mean that the~planet is of a~rocky nature. We~only use this term to make a~connection to the~mass of the~planet we discuss. It~may well be a~that this planet is gaseous, and~the~exact composition does not make a~significant difference to the~applicability of our model. Indeed, a~peculiar composition will reflect in the~observed spectra, but~we expect the~accreting object to have formed from the~very same cloud that formed the~YSO, so their composition should be~similar. 

The~Roche limit for tidal destruction of such a~proto-planet is
\begin{equation}
r_d \approx 2.44 \left( \frac{\rho_1}{\rho_p} \right)^{1/3} R_1
\approx 1.43
\left(\frac{\rho_1}{0.16~\rm{g~cm^{-3}}}\right)^{1/3}
\left(\frac{\rho_p}{1.06~\rm{g~cm^{-3}}}\right)^{-1/3}
\times \left(\frac{R_1}{1.1 \rmRodot}\right) \rmRodot \\ 
\label{eq:r_d}
\end{equation}
where $\rho_1$ is the~density of the~YSO, calculated based on the~parameters derived by~\citep{SiciliaAguilaretal2017}. The~proto-planet is likely to have approached to the~YSO in a~grazing angle, due to an~instability in the~proto-planetary disk. As~Roche limit is larger than the~YSO radius, the~proto-planet would be destructed as~it comes closer to the~YSO by its gravitational field even though its average density is~larger.

Close to the~end of the~$\approx 800 \days$ of accretion, once the~mass supply from the~shredded proto-planet material is terminated, there is an~accretion disk (could be a~thick disk) left that will exhaust itself on a~viscosity timescale.
\begin{equation}
t_{\rm{visc}} \simeq \frac{R_1^2}{\nu} \simeq 55
\left(\frac{\alpha}{0.1}\right)^{-1}
\left(\frac{H/R_a}{0.1}\right)^{-1}
\left(\frac{C_s/v_\phi}{0.1}\right)^{-1}
\times
\left(\frac{R_1}{1.1 \rmRodot}\right)^{3/2}
\left(\frac{M_1}{0.15 \rmModot}\right)^{-1/2} ~\rm{days}, 
\label{eq:tvisc1}
\end{equation}
where $H$ is the~thickness of the~disk, $C_s$ is the~sound speed, $\alpha$ is the~disk viscosity parameter, $\nu=\alpha ~C_s H$ is the~viscosity of the~disk, and~$v_\phi$ is the~Keplerian velocity.
We~get that \textit{$t_{\rm{visc}}$ matches the~time of the~break in the~light curve, discussed above}. This~may indicate that for $\approx2$ months the~light curve is governed (other~than the~contamination from the~rotating spot) by the~depletion of the~accretion disk that the~proto-planet~created. Then comes the~break in the~light curve and~a~shallower decline in brightness.

In~the~above calculation we used a~constant accretion radius of $1.1 \rmRodot$.
Spectral analysis of A13db1417, that found many lines to have inverse P~Cygni or YY~Ori type profiles, lead~\cite{SiciliaAguilaretal2017} to conclude that: (1)~The~YSO and~its proto-planetary disk are viewed almost edge-on. (2)~The~effective temperature during the~eruption is lower than at~quiescence.

We~estimate the~Keplerian time at~the~destruction radius to be $t_k \simeq 0.5 (r_d / 1.4 \rmRodot)^{3/2} \days$. 
The~ratio of viscous to Keplerian timescale is $t_{\rm{visc}}/t_K \simeq 110$. Therefore as~the~accreted material from the~planet is destructed onto the~YSO, it might create a~\textit{thick} accretion~disk.

\begin{figure}[t]
\centering
\includegraphics[trim= 0.0cm 0.0cm 0.0cm 0.0cm,clip=true,width=0.9\textwidth]{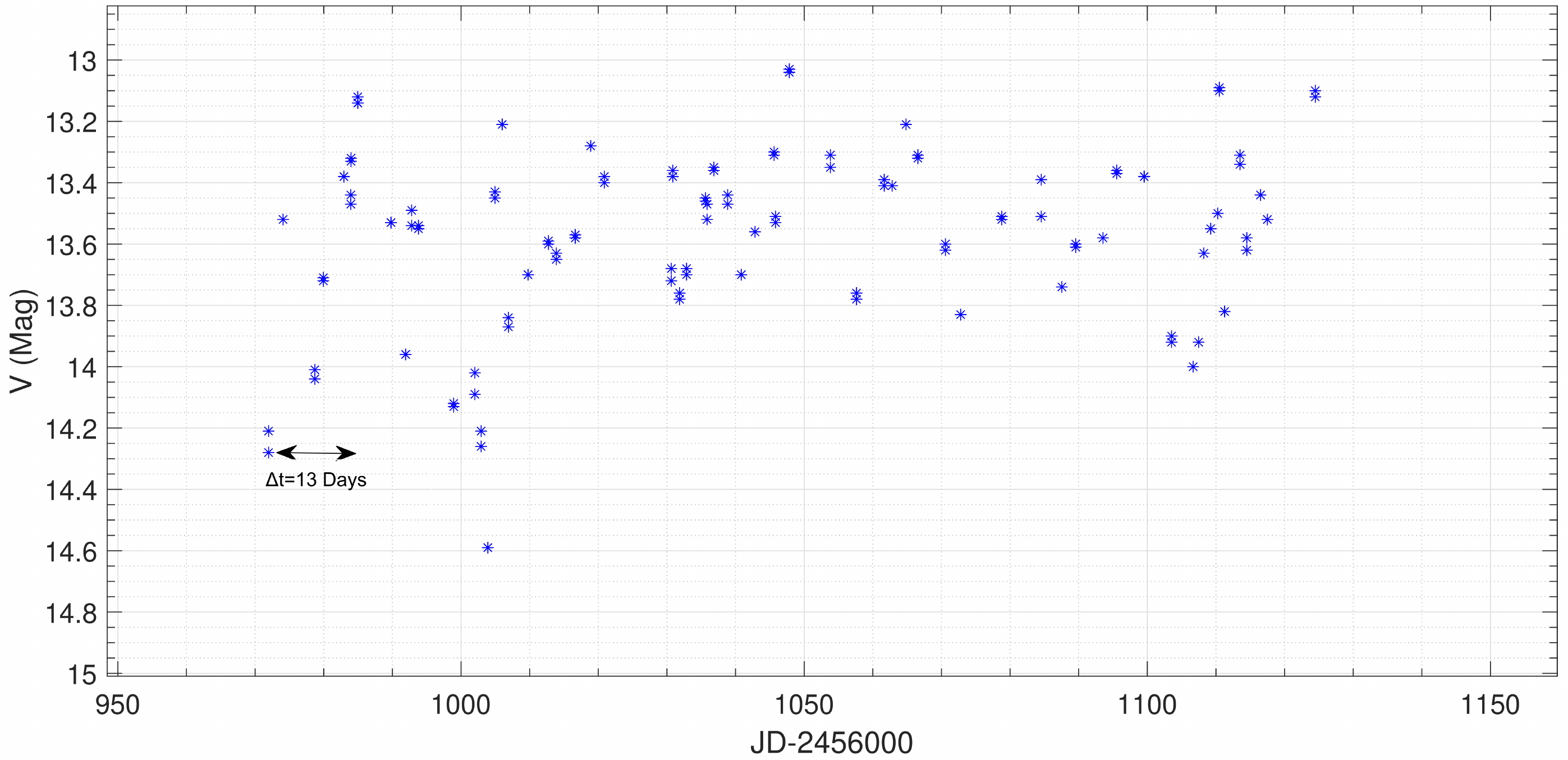} 
\caption{A~focus on the~fast increase in luminosity at~the~beginning of A13db1417.
Observations are taken from~\cite{SiciliaAguilaretal2017}.}
\label{fig:beginning}
\end{figure}

Let us discuss the~beginning of A13db1417. At~the~beginning of the~$\approx 800 \days$ outburst, a~planet is destructed. The~fall-back time for mass to start being accreted is about 6 days for the~parameters we adopted/derived.
Very quickly, in a~matter of a~few dynamical times (namely a~few days) most~of the~destructed material which~falls with a~non-zero angular momentum, will form an~accretion disk. Therefore we expect, and~indeed there is, a~steep increase in the~luminosity. Figure~\ref{fig:beginning} focuses on the~beginning on the~outburst, showing its fast increase in brightness. As~a~characteristic observational constraint to the~increase in the~luminosity at~the~beginning of A13db1417, we~take the~time from the~first detection to the~first local peak, which~about $13 \days$. This~time scale can give a~rough estimate to the~transition time from the~fal-lback accretion, for which~the~destruction time scale is the~relevant timescale, and~the~viscous accretion. We~emphasize that the~source of the~luminosity during the~remainder of the~$\approx 800 \days$ is not from the~destruction itself, but~rather from continuous accretion of the~mass of the~planet onto the~star. It~is therefore possible to very crudely constrain the~fall-back time scale to $\approx 13 \days$, which~is in agreement with the~fall-back time and~the~onset of the~viscous accretion disk.

While the~disk exists it may reduce the~number of ionizing photons escaping from the~YSO close to the~equator, and~also lower the~effective temperature. If~the~YSO, is observed from, or close to, an~edge-on line of sight, this will be evident in the~spectra. Therefore we suggest that the~ILOT model can also account for the~variation in the~observed line~intensities.

\begin{figure}[t]
\centering
\includegraphics[trim= 0.0cm 0.0cm 0.0cm 0.0cm,clip=true,width=0.9\textwidth]{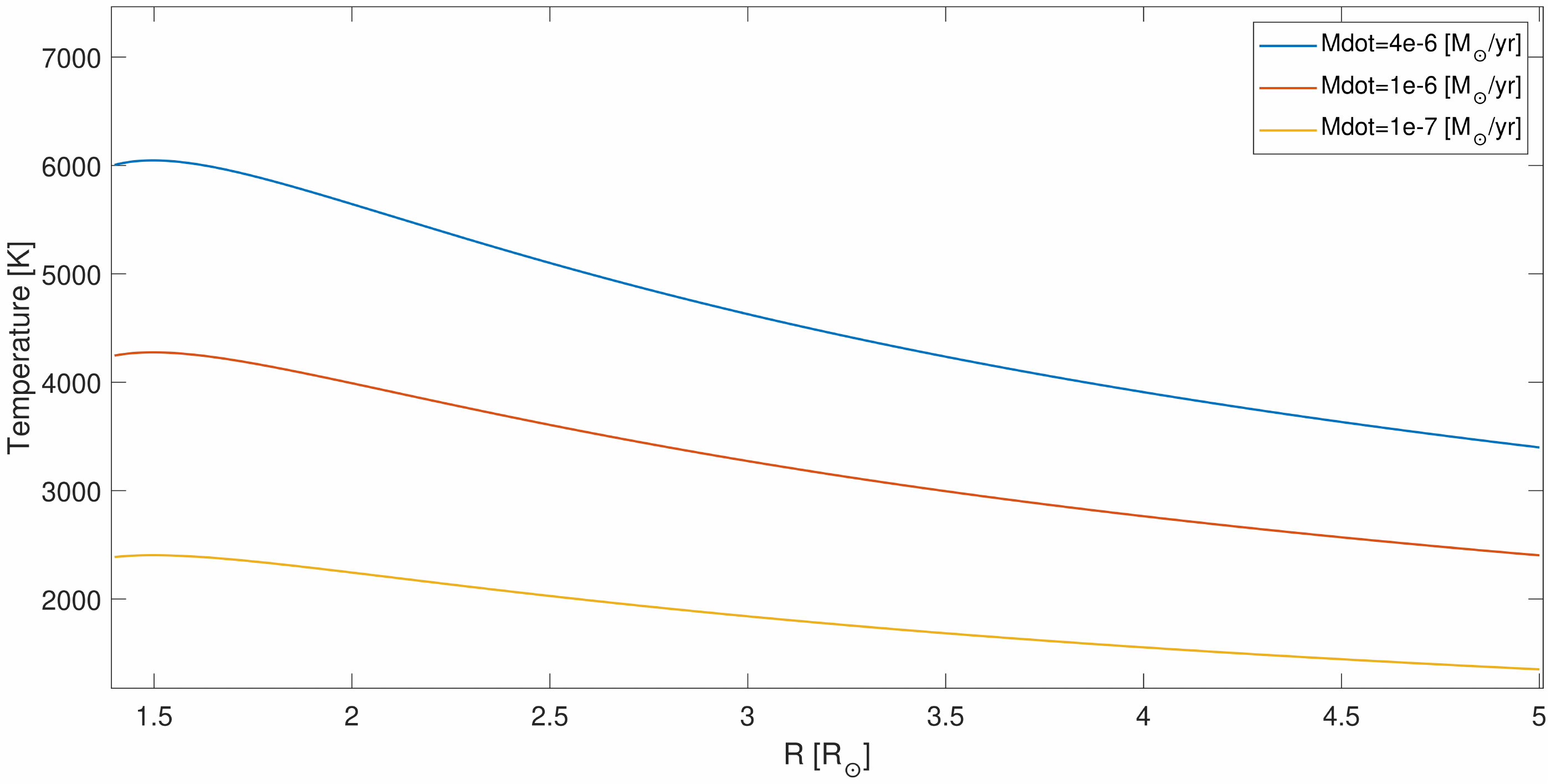}
\caption{An~estimate to a~classical accretion disk surface temperature, according to Equation~(\ref{eq:TR}). When~the~disk is depleted the~mass accretion rate decreases and~so does the~surface temperature of the~disk.}
\label{fig:TR}
\end{figure}

Analyzing the~data, we find that the~effective temperature of A13db1417 declined from $\simeq$$4500\K$ to $\simeq$$2000\K$ during the~last stage of the~eruption. The~trend and~the~values are similar to the~effective temperature from of V838~Mon~\citep{Tylenda2005}. According~to our model, the~emission of A13db1417 comes from a~viscous disk. Modeling the~emission in our specific case would require an~extensive numerical study, that might be too excessive given the~rough estimate of the~parameters. But~for a~crude estimate, we~calculate the~temperature for a~classical accretion disk~model
\begin{equation}
T(R)=\left( \frac{3 G M_1 \dot{M}_{\rm acc} }{8 \pi R^3 \sigma} \left[1-\left(\frac{R_1}{R}\right)^{1/2}\right] \right)^{1/4}
\label{eq:TR}
\end{equation}
where $R$ is the~disk radius and~$\sigma$ is the~Stefan-Boltzmann constant. Figure~\ref{fig:Tt} shows the~surface temperature for the~disk according to Equation~(\ref{eq:TR}). As~the~disk depletes during \vspace{2pt} the~$\approx 55 \days$, the~mass accretion rate~drops. Consequently, the~temperature drops as~$T \sim \dot{M}_{\rm acc}^{1/4}$.
The~outer part of the~disk, that has the~largest surface area and~thus dominant in the~spectrum, has a~surface temperature in the~range of $\simeq3500$--$4500\K$. During the~decline in the~emission of the~outburst drops by $\Delta V \simeq 4$~mag in $\approx 55 \days$ (see~Figure~\ref{fig:v838_VS_13db}), corresponding to a~drop in the~brightness by a~factor of $\simeq40$. Therefore, according to the~model, $\dot{M}$ needs to drop by that factor as~well. As~the~accretion rate drops by that factor of 40 the~temperature drops by a~factor of $\simeq2.5$. We~note that Equation~(\ref{eq:TR}) describes a~steady state and~does not accurately account for the~thermal changes during the~dynamic phase of the~depletion. Given the~crude estimate, it turns out that the~disk surface temperature is consistent with the~observations that show a~surface temperature drop to $\approx2000\K$ at~the~end of the~outburst (Figure~\ref{fig:TR}).

%

\section{Summary and~Discussion} 
\label{sec:summary}

We~discussed the~eruptions of the~YSO ASASSN-13db, which~erupted twice: a~short eruption in~2013, and~a~longer one from~2014 to~2017. The~2013 event is a~typical YSO outburst, but~the~2014--2017 event (A13db1417) is clearly different. The~object may be intermediate between FUor and~EXor outbursts. It~has been proposed that there is a~continuous spectrum of variables between EXors and~FUors~\citep{ContrerasPenaetal2017}, based on the~outburst ranges, so~that some bursts may be more FUor-like and~some others more EXor~like.

We~found a~strong resemblance in the~decline light curves of A13db1417 and~V838~Mon, a~prototype LRN ILOT. The~two outburst have similar decline in their light curves, which~we suggest may not be a~coincidence. Namely, we follow~\cite{Kashietal2010} who suggested that mass accretion that releases energy in a~specific rate can account for a~special shape of light curve, that will be similar for different objects when rescaled. There~is also a~similar decline in the~effective temperature of the~two transients. The~evolution of the~temperature also matches the~expected decline of the~surface temperature of a~classical accretion disk, as~a~result from a~decrease in the~accretion rate. Therefore we examine the~possibility that A13db1417 is an~accretion-powered event, as~was suggested for V838~Mon. We~found that the~time from the~peak of the~light curve of A13db1417, to~the~break were it begins a~much shallower decline is the~same as~the~viscosity time on which~the~shredded proto-planet is accreted, and~therefore the~supply of mass for gravitational energy to power the~event is~consumed.

Most stars are expected to form planets at~some point, and~during this process the~orbits are in many cases unstable, what may lead to a~collision course between the~proto-planet and~the~proto-star. We~examined the~possibility that A13db1417 was caused by accretion of a~proto-planet onto the~proto-star, in what is termed an~Intermediate Luminosity Optical Transient~(ILOT).

It~is possible that the~A13db1417 eruption is accretion not from a~destructed planet but~rather of gas from the~proto-planetary disk. The~amount of accreted mass in this case would be the~same as~calculated in Section~\ref{sec:model}, from the~energy considerations. In~that case there should be an~instability in the~proto-planetary disk, that would increase the~normal mass accretion rate to a~few $\times 10^{-6} \msyr$ (see~Equation~(\ref{eq:mdotacc})). If~ASASSN-13db will have more eruptions with the~same declining slope it will be in favor of the~instability in the~proto-planetary disk scenario, as~it is less likely that a~number of planets will interact with the~YSO in the~same manner. The~proto-planetary disk, on the~other hand, does not have such a~constraint, and~multiple instabilities that may lead to eruptions are~possible.

In~Section~\ref{sec:model} we estimated that the~total energy involved in the~A13db1417 eruption is $ \approx 10^{42} \erg$. If~the~outburst is indeed an~ILOT, A13db1417 is the~least energetic one observed to date. The~Eddington ratio for A13db1417 is $\Gamma_{\rm Edd} \simeq 10^{-4}$, which~is very small compared to objects in the~OTS, that can reach even $\Gamma_{\rm Edd} \simeq 10$. 
It~is therefore located at~the~bottom of the~Energy Time Diagram (ETD), as~a~very weak eruption in terms of energy (as~low as~the~energy of novae and~below), and~an~intermediate duration in the~scale of~months.

\textls[-10]{ASASSN-13db is an~important contribution to the~ILOT family, as~it expands the~region on the~ETD where ILOTs can reside. We~see~that an~ILOT can be not ``intermediate'', in~the~sense that its luminosity does not necessarily be in the~gap between novae and~SNe. The~word ILOT does not classify an~object in terms of its brightness, but~rather designates an~object with a~specific physical process, accretion~and~release of gravitational~energy.}

The~light curve of V838~Mon itself, which~was used here as~a~reference is not yet completely understood. It~was suggested to be the~result of the~merger-burst process, and~was qualitatively obtained from analytic considerations, but~further investigation is~required. An~different interpretation of the~light curve of the~prototype LRN, V838~Mon, as~a~sequence of three planet collision with the~a red giant was proposed by~\cite{RetterMarom2003} and~\cite{Retteretal2006}. Setting aside the~fact that the~progenitor of V838~Mon was not a~red giant, the~scenario they proposed could not account for the~energy of the~outburst, and~it is nearly impossible that three planets would fall one after another onto their parent star on a~timescale of months~\citep{TylendaSoker2006}. Our~scenario is different than the~scenario proposed by~\cite{RetterMarom2003} as~the~planet does not have to arrive in a~direct collision course, and~it does not collide with the~star but~rather shredded before it~arrives.

The~model we consider here for the~2014--2017 eruption of ASASSN-13db (A13db1417), even~though it involves a~planet, has none of the~issues listed above. It~is a~single event, that happened unrelated to the~2013 eruption of ASASSN-13db. Also, the~model is designed specifically for the~energy budget of~A13db1417.

\textls[-15]{A13db1417 is another example of the~process that manifest itself with the~characteristic ILOT~lightcurve. We~postulate that this process is the~depletion of an~accretion disk. To~show that, 3D~numerical simulations are required, together with radiation-transfer analysis.}
This~remains to be studied in a~future~work. 
\vspace{6pt}

\authorcontributions{All authors have read and~agree to the~published version of the~manuscript.} 

\funding{This~research received no external funding.} 

\acknowledgments{We~acknowledge support from the~R\&D authority, and~the~Chairman of the~Department of Physics in Ariel University.}

\conflictsofinterest{The~authors declare no conflict of interest.} 
\reftitle{References}

\end{document}